\documentclass[aps,prl,showpacs,floatfix,twocolumn]{revtex4}
\usepackage{amssymb}
\usepackage{amsmath}
\usepackage{graphicx}
\begin{document}
\title{The single-atom box: bosonic staircase and effects of parity}

\author{D.V. Averin}
\affiliation{Department of Physics and Astronomy, University of
Stony Brook, SUNY, Stony Brook, NY 11794-3800}
\author{T. Bergeman}
\affiliation{Department of Physics and Astronomy, University of
Stony Brook, SUNY, Stony Brook, NY 11794-3800}
\author{P.R. Hosur}
\affiliation{Department of Physics, Indian Institute of Technology Bombay,
Mumbai 400076, India}
\author{C. Bruder}
\affiliation{Department of Physics, University of Basel,
CH-4056 Basel, Switzerland}


%
%

\begin{abstract}
We have developed a theory of a Josephson junction formed by two
tunnel-coupled Bose-Einstein condensates in a double-well potential
in the regime of strong atom-atom interaction for an arbitrary total
number $N$ of bosons in the condensates. The tunnel resonances in
the junction are shown to be periodically spaced by the interaction
energy, forming a single-atom staircase sensitive to the parity of
$N$ even for large $N$. One of the manifestations of the staircase
structure is the periodic modulation with the bias energy of the
visibility of the interference pattern in lattices of junctions.
\end{abstract}

\maketitle

Bose-Einstein condensates (BECs) have a unique ability to give rise
to individual quantum states occupied with a macroscopically large
number of particles. The interference of the wavefunctions of such
macroscopic states leads to a variety of macroscopic quantum
phenomena, the most studied of which is the Josephson effect: the
flow of current across a potential barrier due to the interference
of the condensate states on the two sides of the barrier. For a long
time since the original discovery \cite{jj}, this effect has been
studied in solid-state ``Josephson junctions'' \cite{kkl}: two
weakly-coupled superconductors, in which the condensates are formed by
Cooper pairs \cite{tnk}. At energies
smaller than the binding energy of electrons in Cooper pairs, the
characteristics of Josephson junctions can be discussed directly in
terms of the properties of the Cooper-pair condensates \cite{dva}.
In this case, the two dynamic variables of the Josephson junction
(JJ) are the phase difference $\phi$ between the condensate
wavefunctions on the two sides of the barrier and the
half-difference $n$ of the number of Cooper pairs in the two
condensates. Since the phase $\phi$ and the ``charge'' $n$ are
conjugate quantum variables \cite{and}, the dominant qualitative
feature of the JJ dynamics is the transition from the classical dynamics
of $\phi$ in large junctions to the classical dynamics of $n$ in
small junctions. The classical charge regime is characterized by the
correlated transfer of individual Cooper pairs \cite{azl}, with the
simplest ``Cooper-pair box'' system \cite{mb,sac} providing the most
basic demonstration of this transfer in the ``Cooper-pair
staircase'': the step-wise increase of $n$ with the junction bias
that corresponds to successive transfers of Cooper pairs one by one
through the junction. The transition between the quasiclassical
dynamics of $\phi$ and $n$ is affected by the energy dissipation in
the junction \cite{sch} (with the classical-$n$ regime requiring
small dissipation) but can generally be understood as competition
between the tunnel amplitude and the interaction energy of Cooper
pairs.

In the case of solid-state JJs, the bosonic junction model is partly
a pedagogical simplification. With ultracold atoms in optical traps
of double-well structure, this model, however, can
\cite{milburn,smerzi} and recently has been realized directly. It
has been studied experimentally both in the regime of
larger-scale BECs in the wells with relatively weak transverse
confinement \cite{ajj1,ketterle,ajj2}, when they exhibit the
Josephson effect governed by the classical dynamics of the phase
difference $\phi$, and in the ``Mott insulator'' regime
\cite{bl1,win}, when the total number $N$ of atoms in the
double-well structure is small $N=1,2$ and the number difference $n$
behaves quasiclassically. One of the main dynamic features of the
regime of classical $n$ is the existence of resonances, as a
function of the energy bias $\epsilon$, corresponding to the
tunneling of individual atoms \cite{bl2,schmelcher,lee2008}. The goal of
this work is to study these ``single-atom'' resonances for an
arbitrary number of atoms $N$ in the BEC Josephson junction, in both
the interaction-dominated Mott-insulator regime and the case of
large-scale BECs. We show that the main features of the single-atom
effects in BEC junctions are quite similar to those in solid-state
JJs \cite{azl,mb,sac}, and in particular, include the periodic
spacing of resonances in the ``single-atom staircase''. There are,
however, several new effects produced by the fixed total number of
bosons $N$ in the BEC junction, as opposed to solid-state JJs where
the total number of particles is not well-defined. One is the
modulation of the strength of the resonances with the number
difference $n$, which reflects the collective nature of the tunnel
coupling of the BECs even in the regime of single-atom tunneling.
Another is the dependence of the resonances on the parity (even or
odd) of the total number of bosons $N$.

We consider $N$ bosons in a double-well potential approximated by a
two-site Bose-Hubbard model \cite{jbc,foerster,carr,salgueiro}:
\begin{equation}\label{h1}
H_0 = \frac{U}{2}\sum_{i=1,2} n_i(n_i-1) -\Delta (a_1^\dagger a_2 +
a_2^\dagger a_1) - \epsilon (n_1-n_2) \, .
\end{equation}
Here $n_i=a_i^\dagger a_i$ and $a_i^\dagger$ ($a_i$) creates
(annihilates) a boson in the lowest-energy state localized in the
$i$th well, $\Delta$ is the tunnel coupling of these two states,
$2\epsilon $ is their difference in on-site energy, and $U$ is the
on-site interaction. The Hamiltonian (\ref{h1}) conserves the total
number of bosons $N=n_1+n_2$, and after separation of $N$, it takes
the following form (up to a constant energy shift):
\begin{equation} \label{h2}
H_0 = U n^2 - 2 \epsilon n - \Delta (a_1^\dagger a_2 + a_2^\dagger
a_1) \, .
\end{equation}
Here $n \equiv (n_1-n_2)/2$ is the occupation number difference
between the two wells. While $N$ does not enter the Hamiltonian
(\ref{h2}) explicitly, it influences the junction dynamics by limiting
the range of the possible values of the number difference 
$n \in \{-N/2, -N/2+1, ... , N/2\}$, i.e., to integer or half-integer $n$ for
even or odd $N$, respectively. In general, the kinematics of the model
defined by Eq.~(\ref{h1}) or (\ref{h2}) gives Schwinger's
representation of the angular momentum $\vec{J}$ of magnitude $N/2$
(see, e.g., \cite{mat}). In this identification, $n$ is the
$z$-component $J_z$ of the momentum, and the tunneling term in the
Hamiltonian is $-2 \Delta J_x$. Since the properties of the integer
and half-integer momenta are quite different, this {\em parity effect}
makes the energy characteristics of the bosonic junction (\ref{h2})
dependent on the parity of $N$, as we show in more detail below.
\begin{figure}[htb]
\includegraphics[width=8.2cm,clip=true]{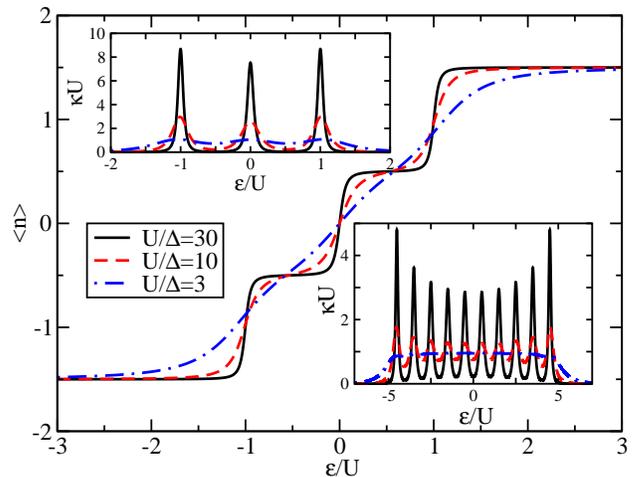}
\caption{Average number difference $\langle n\rangle $ in the ground
state of the double-well BEC Josephson junction for $N=3$ as a
function of the energy difference $\epsilon$ between the two wells,
for several values of the ratio of interaction energy versus tunnel
amplitude $U/\Delta$. Curves with $U/(N \Delta) \gg 1$ show the
single-atom staircase, which is smeared with increasing tunneling
strength. Left inset: Compressibility $\kappa=\partial \langle n
\rangle/\partial\epsilon$ showing the peak structure associated with
the single-atom resonances. For odd $N$, there is a resonance in the
unbiased junction $\epsilon \simeq 0$.
Right inset:
$\kappa$ for $N=10$ bosons in the two
wells. With increasing tunneling strength, $\kappa$ changes from
sharp single-atom resonances to a quasiclassical behavior with
constant large-scale compressibility $\simeq 1/U$ and weak
sinusoidal single-atom oscillations.} \label{fig1} \end{figure}

To see this quantitatively, we start with the regime of strong
interaction, $U\gg N \Delta $. In this case, the tunnel coupling
$\lambda \equiv \Delta [(N/2+n+1)(N/2-n)]^{1/2}$ of the states $n$
and $n+1$ that differ by the transfer of one atom in the junction,
is small, $\lambda \ll U$. This means that these states are coupled
effectively only in the vicinity of certain values of the energy
bias $\epsilon\approx\epsilon_n$,
\begin{equation} \label{res}
\epsilon_n \equiv (n+1/2)U\;,
\end{equation}
when the interaction energy difference between them is suppressed
and atoms can tunnel across the junction. Since these resonances are
uniformly spaced by $U$, the average number difference $\langle
n\rangle$ in the ground state of the junction shows a regular {\em
staircase-like structure} (Fig.~\ref{fig1}) similar to the
corresponding structure in the Cooper-pair box \cite{mb,sac}. The
curves in Fig.~\ref{fig1} are obtained by straightforward numerical
diagonalization of the Hamiltonian (\ref{h2}) and show also
the smearing of the staircase by tunneling when it is increased to
the level $\Delta \sim U/N$. At $U \gg N\Delta$, the system dynamics
reduces in the vicinity of each step to the two states $n$ and
$n+1$, and one can find directly the shape of the steps in the
staircase:
\begin{equation} \label{st}
\langle n\rangle = n+\frac{1}{2}\left(1+ \frac{\epsilon-\epsilon_n}{
[(\epsilon -\epsilon_n)^2+\lambda^2]^{1/2} }\right) \, .
\end{equation}

The shape of each of the single-atom transitions is seen more clearly
in the resonant peaks in the compressibility $\kappa = \partial
\langle n\rangle/\partial\epsilon$ of the ground state. In particular,
Eq.~(\ref{st}) shows that the maximum of each peak reflects directly
the effective tunnel coupling of the two states:
\begin{equation} \label{com}
\kappa = \frac{1}{2\lambda} =\frac{1}{2\Delta} [(N/2+n+1)
(N/2-n)]^{-1/2} \, .
\end{equation}
This equation agrees with the curve for strongest interaction in the
right inset to Fig.~\ref{fig1}, and shows that the enhancement of the
tunneling amplitude from the single-particle value $\Delta$ to the
multi-particle value $\lambda \sim N \Delta$ by the BEC coherence
manifests itself even in the regime of single-atom tunneling. The
modulation (\ref{com}) is a quantum consequence of the dependence of
the critical current of the BEC junction on the population imbalance
between the two wells that also leads to ``self-trapping'' of the
condensate in the case of classical Josephson dynamics
\cite{ajj1,ajj2,ab}.

The main qualitative feature of the staircase that reflects the
parity effect is that the steps (\ref{st}) occur at half-integer
values of energy bias $\epsilon$ (in units of interaction energy
$U$) for even $N$ and integer $\epsilon/U$ for odd $N$, see
Eq.~(\ref{res}). In particular, the tunneling is always suppressed
for even $N$ in the symmetric unbiased junction, when $\epsilon
\simeq 0$, but there is a tunneling resonance at $\epsilon=0$ for
odd $N$ (Fig.~\ref{fig1}). This resonance at $\epsilon=0$ has an
obvious interpretation in the single-atom regime with strong
interaction energy, $U\gg N\Delta$: the interaction energy of $N$
atoms is degenerate with respect to the last atom out of an odd
total number $N$ being placed in the left or right well. This simple
interpretation is not valid in junctions with large $N$, when $U\ll
N\Delta$, but the parity effect exists even in this case, provided
the interaction energy still plays a role in the junction dynamics,
$U\gg \Delta/N$. (If $U\ll \Delta/N$, the interaction is not
essential and the junction represents a collection of $N$
independent two-state systems of individual atoms.) In this regime,
the number-difference fluctuations $\delta n$ around the average
value $\langle n \rangle=\epsilon/U$ have the magnitude (estimated
from the Hamiltonian (\ref{h3}) below) $1 \ll \delta n \simeq
(N\Delta/U)^{1/4} \ll \sqrt{N}$. If $\langle n \rangle$ is not too
close to the boundaries $\pm N/2$ of the allowed interval for $n$, this
means that effectively there are no restrictions on the values of $n$, and
variations of the tunneling amplitude $\lambda$ are negligible. As a
result, one can define a phase $\phi \in[0,2\pi] $ conjugate to $n$
and the Hamiltonian (\ref{h2}) takes a form that essentially
coincides with that of the ``solid-state'' Josephson junction:
\begin{equation} \label{h3}
H = -\frac{\epsilon^2}{U}- U\frac{\partial^2}{\partial \phi^2} -2
\lambda  (\epsilon) \cos \phi\, ,
\end{equation}
where $\lambda (\epsilon)= \Delta [(N/2)^2-(\epsilon/U)^2]^{1/2}$.
The boundary conditions on the wavefunctions $\psi (\phi)$, however,
depend on the bias $\epsilon$ and the parity of $N$:
\begin{equation} \label{bc}
\psi (\phi+2\pi) =(-1)^N e^{-i2\pi \epsilon/U}\psi (\phi) \, .
\end{equation}
Using the known result \cite{azl} for the Hamiltonian (\ref{h3}) we
see that the ground state energy of the bosonic junction is
\begin{equation} \label{gr}
E_0= -(\epsilon^2/U) - (-1)^N \delta_0 \cos(2\pi \epsilon/U) \, ,
\end{equation}
where $\delta_0\simeq (1/\pi)(\lambda U)^{1/2}
\exp[-8(\lambda/U)^{1/2}]$. The first term in (\ref{gr}) describes
the large-scale, mean-field compressibility $\kappa=-(1/2)
\partial^2 E_0/\partial \epsilon^2= 1/U$ for $\epsilon$ within the
range $[-NU/2,NU/2]$ (right inset to Fig.~\ref{fig1}).
Qualitatively, this compressibility is dominated by the interaction,
since the total interaction energy scales as $N^2$ as opposed to the
tunneling energy which is proportional to $N$. The second term in
(\ref{gr}) describes the sinusoidal single-atom oscillations of
small amplitude which increase at the ends of the range of
$\epsilon$ because of the decrease of $\lambda$ (right inset to
Fig.~\ref{fig1}). These oscillations represent what is left of the
single-atom resonances when they are washed out by strong tunneling,
i.e., describe the discreteness of the BEC flow in this
quasiclassical regime. The position of the oscillations as a
function of the bias $\epsilon$ depends on the parity of $N$, with a
minimum in $\kappa$ at $\epsilon\simeq 0$ for even $N$ and a maximum
for odd $N$.

The sensitivity of the bosonic JJ to individual atoms makes it
interesting to consider the situation with one extra ``foreign''
atom in the junction. Motivated by experiments on Bose-Fermi
mixtures \cite{esslinger2006}, we assume that this extra particle is
a ``fermion'', although its statistics does not play a role in our
discussion. Besides the bosonic part $H_0$ (\ref{h1}), the junction
Hamiltonian includes then a fermionic part $H_f$ of the same
structure with the tunnel amplitude $\Delta_f$, the number operator
$n_{fi}$, on-site interaction $U_f$, and a boson-fermion interaction
term:
\begin{equation}\label{ham}
H=H_0+H_f+ U_{bf}(n_1 n_{f1}+n_2 n_{f2})\; ,
\end{equation}
The boson-fermion interaction energy $U_{bf}$ can be positive or negative
\cite{esslinger2006}. The difference $\epsilon$ in on-site energies
is assumed to be the same for bosons and the fermion.

\begin{figure}[htb]
\includegraphics[width=7.5cm,clip=true]{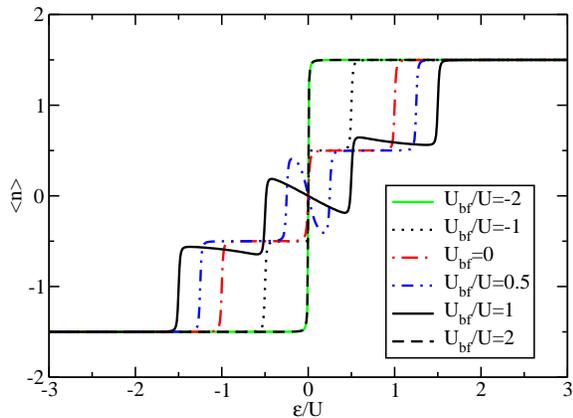}
\caption{(color online) Steps in $\langle n\rangle$ as a function of
$\epsilon$ for $N=3$ bosons, $N_f=1$ fermions, $\Delta/U=0.01$, and
$\Delta_f/U=0.004$.} \label{figure_steps_fermion}
\end{figure}

One can follow separately the dynamics of the two types of
particles, and we limit our brief discussion to the behavior of the
bosons. The average boson number difference $\langle n \rangle$ is
shown in Fig.~\ref{figure_steps_fermion} for $N=3$. In the case of
vanishing boson-fermion interaction, $U_{bf}=0$, we obtain the
familiar three steps of height 1. For strong attraction, $U_{bf}\ll
-U$, or repulsion, $U_{bf}\gg U$, minimization of the interaction
energy requires that the bosons are moving all together, and
$\langle n \rangle$ shows only one step of height 3 at $\epsilon=0$.
In our example of 3 bosons, this case is realized already for
$|U_{bf}| \geq 2U$. In the intermediate region $|U_{bf}| \leq 2U$,
we find a more complex behavior: for $U_{bf}<0$ and $U_{bf}>0$ the
bosonic steps are shifted, respectively, to and away from the region
$\epsilon=0$. Moreover, for $U_{bf}>0$ (i.e., for antiferromagnetic
coupling between the angular momenta representing bosonic and
fermionic parts of the junction), the system is frustrated and the
steps have regions with negative slope, see the curves for
$U_{bf}/U=0.5$ and $1$. In the ``single-atom'' regime of small
tunneling amplitudes, and for odd $N$, the region of the negative
slope around $\epsilon \simeq 0$ corresponds to the range $\epsilon
\in [-U_{bf}/2,U_{bf}/2]$, where $\langle n \rangle$ is described by
the following general relation:
\begin{equation} \label{bf}
\langle n\rangle = \frac{1}{2}\frac{\epsilon (\Delta_f^2 -\lambda^2)
}{ [\epsilon^2 (\Delta_f^2 -\lambda^2)^2 +\lambda^2\Delta_f^2
U_{bf}^2]^{1/2} } \, .
\end{equation}
Note that the slope depends of the relation between
the effective boson and fermion tunneling amplitudes.

\begin{figure}[htbp]
  \centering
  \includegraphics[width=7.5cm, clip=true]{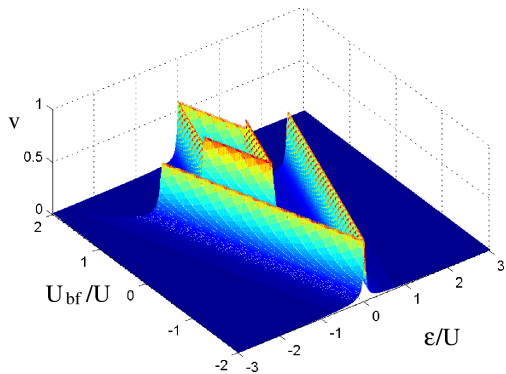}
\caption{(color online) Visibility $v$ of the bosonic interference
  pattern in the Bose-Fermi junction. All the parameters are chosen as
  in Fig.~\protect\ref{figure_steps_fermion}.} \label{visibility}
\end{figure}

One way of studying the effects discussed in this work would be to
measure the interference pattern of the expanding atomic cloud after
switching off an optical lattice of appropriately spaced
uncoupled double-well junctions. Such arrays of double wells have
already been realized experimentally \cite{porto}. The interference
of particles within each junction will produce a cosine modulation
of the average atom intensity $I$. The visibility of this modulation
pattern is given \cite{interf} by the average kinetic (tunneling)
energy per boson in units of $\Delta$:
\begin{equation*}
v=(I_{max}-I_{min})/(I_{max}+I_{min})= \langle a_1^\dagger a_2 +
a_2^\dagger a_1 \rangle/N\, .
\end{equation*}
In the single-atom regime of the bosonic junction (\ref{h2}), the
visibility is modulated by the bias $\epsilon$ and reaches its
maximum at the resonances (\ref{st})
\begin{equation} \label{vis}
v =(\lambda/N)[(\epsilon -\epsilon_n)^2+\lambda^2]^{-1/2} \, ,
\end{equation}
where the states of the two wells deviate most strongly from pure
number states. For the Bose-Fermi junction (\ref{ham}), $v$ shows
a rich dependence on $\epsilon$ and
$U_{bf}$ that is also related to the staircase structure, see
Fig.~\ref{visibility}.

In conclusion, we have shown that a Josephson junction formed by two
tunnel-coupled BECs shows single-atom effects similar to those in
solid-state Josephson junctions.  In addition to periodically spaced
resonances as a function of the difference in on-site energy leading
to a single-atom staircase, there are modulations of the strength of
the resonances and parity effects that can be traced back to the
fixed total number of bosons $N$ in the BEC junction. A different,
e.g. fermionic, additional particle in the junction leads to
non-trivial modifications of the staircase, that can be experimentally
observed in the visibility of the interference pattern.

\begin{acknowledgments}
We thank D. Schneble for discussions. During the preparation of this
manuscript, we became aware of a similar work by G. Ferrini 
{\it et   al.} \cite{ferrini}. Very recently, the periodic spacing of the
single-atom resonances discussed in this work was observed
experimentally \cite{bloch2008} using a different detection scheme.
Our work was supported in part by the US NSF grants \# DMR-0325551 and
PHY-0652459, the Swiss NSF, the NCCR Nanoscience, and the European
Science Foundation (QUDEDIS network).
\end{acknowledgments}

\end{document}